# GLOBAL SURFACE WAVE RESONANCES OF THE EARTH'S MAGNETOSPHERE AND THEIR POSSIBLE MANIFESTATION

Petko Nenovski

University Center for Space Research and Technologies (UCSRT), Sofia University "St. Kliment Ohridski", Sofia, Bulgaria

**Abstract**

In this paper global surface wave (SW) modes supported by plasma discontinuities at both the magnetopause and the plasmapause are considered. The ionosphere at the ends of the magnetic field lines of the outer magnetosphere is considered as reflecting boundaries of the SW that propagate along the plasma. As a result a standing wave structure along the magnetic field fluxes of the outer Earth's magnetosphere (surface wave resonance (SWR)) can be formed. Due to quantized wavenumbers along the magnetic field lines, the SWR possesses quantified frequencies in a following way: $f_{1,2} \equiv (1, 2, 3,..)f_{0,1(2)}$, where $f_{0,1(2)}$ is the frequency of the corresponding fundamental SWR.

Global Pc5 pulsations have been observed and interpreted mostly as cavity modes of the Earth's magnetosphere**.** The global Pc5 pulsations of discrete spectral peaks however could alternatively be interpreted as low-frequency surface wave resonances (SWR) of the Earth's magnetosphere that do not necessarily involve the cavity mode–field-aligned resonance (FLR) transformation concept.

**Introduction.** ULF geomagnetic variations (1 mHz÷10 Hz) were first observed in the ground-based measurements of the 1859 Carrington Event [1]. ULF variations are divided into regular (Pc) and irregular (Pi) pulsations and subdivided in frequency [2]. Nowadays ULF geomagnetic variations are considered as a manifestation of magnetohydrodynamic (MHD) waves or electric currents structures emerging in the Earth's magnetosphere. Best known MHD waves in the Earth's magnetosphere are Alfven, magnetosonic, field-aligned resonances (FLR), cavity/waveguide modes, etc. One branch of regular ULF pulsations – the Pc5 pulsation events, have period 150÷600 s and are observed mostly as FLR at high latitudes. Occasionally Pc5 pulsations however emerge in the Earth's magnetosphere both at high, mid- and low latitudes and then are called global Pc5 pulsations. The latter are long-lived structures with discrete frequencies and sometimes have large amplitudes. Currently they are considered as a manifestation of cavity/waveguide MHD modes. Let us consider this hypothesis in detail.

**Little preface**. First pioneering observations of discrete modes of frequencies 1.3, 1.9, 2.6, 3.4 mHz in the Pc5 pulsation range were performed by the Goose Bay radar [3-5]. Lately, similar observations from the Wick radar at mid latitudes [6] have provided independent evidence of similar discrete field line resonances (FLR). Thus, three distinct spectral peaks with frequencies of 1.8 mHz, a broad region of enhanced spectral power at 2.4-2.8 mHz and a peak at 3.2 mHz have been observed at mid latitudes [6]. The important feature is namely an existence of distinct "hot-spots" in frequency-latitude space, where enhanced ULF wave activity mostly is concentrated. A comparison of spectra observed by the Goose Bay and Wick radar has demonstrated that the frequencies of the observed modes are, on average, almost identical, despite the different latitudinal bands covered by the two radar systems [6]. These Pc5 pulsations represent long duration events under high speed solar wind conditions. Similar long-duration Pc5 events and high and low latitudes have been observed by Francia and Villante [7], Villante et al [8,9]. Recently, Marin et al [10] have studied longitudinal and latitudinal properties of global Pc5 pulsation activities during strong magnetic storms occurred on 28-31 October 2003. Field-line resonances (FLRs) of such low frequencies (< 4 mHz) cannot exist at mid and low latitudes. Hence, the latter cannot be considered as possible mechanisms of the global Pc5 pulsations. The observed low frequency peaks of the global Pc5 pulsations are thus attributed to so-called cavity modes [11-15]. The discrete frequency set of the cavity modes then comes from the distance $\Delta R$ (in equatorial plane) between the magnetopause (or bow shock) and the frequency-dependent turning point at the plasmapause. The frequency of waveguide modes is determined by the quantization condition $k\Delta R \sim \pi n$, where $k$ is the wavenumber, and $n$ is an integer. Obviously, their wave field structure is oscillatory in the outer magnetosphere and decays exponentially beyond the turning point (near the plasmapause). According to the theory, the cavity mode observation on the ground becomes possible due to field line resonances (FLR) excited only at latitudes where they match the eigenfrequency of standing wave structure formed on the corresponding field line [15]. Thus, 1.8 mHz FLR can exist only at high latitudes (approximately at TNB latitudes [8,9]) where its magnetic field should be oriented in N-S direction. The polarization picture is however inverse one – dominance of the D component at high latitudes (TNB), while a prevalence of the H component is registered at low latitude station (L'Aquila). In addition, the standing wave structures as expected from cavity mode excitation mechanism are of high frequencies (usually above 10 mHz) [14]. In order to fit the observed low-frequency (2.7 mHz) of the global Pc5 mode to cavity mode mechanism even a unlikely fundamental quarter-wavelength wave structure has been suggested [10]. The existence of mode of constant, low-frequencies under strongly variable magnetopause position conditions is the main obstacle that cannot be reconciled by the cavity mode theory. In general, the experimental facts cannot be explained in the frame of the cavity theory, although some of them (e.g. global Pc5 pulsations of frequency 1.4÷1.6 mHz have been interpreted as waveguide mode [e.g. 5, 16].

The observed Pc5 pulsations and their properties, i.e. that are of global characteristics and giant (sometimes) amplitudes as well as, the lack of well justified explanation of their amplitude, discrete spectrum belonging to the Pc5 range, phase and polarization properties, are the main motivation of this paper. This study makes an important point about surface wave (SW) properties that has been overlooked in past: SW are not necessarily strongly localized at the magnetopause boundary and therefore, omitted as possible mechanism of global Pc5 pulsation events. The first goal of the present study is to enlighten those SW mode properties (global in extent, discrete frequencies, etc.) that are of relevance to observations of global Pc5 pulsations and their mechanisms. The second goal is to consider the similarities and difference between the global surface wave mode and the field-aligned resonances (FLR). The third goal is to demonstrate that SW theory can explain many observational features that cavity waveguide modes fail to do.

**The surface wave theory. Basic properties of SW resonances (SWR).** With a few exceptions, no attention has been paid to another class of possible global modes in the Earth's magnetosphere: Alongside with surface waves (SW) supported by one boundary or Kelvin-Helmholtz instability (on magnetopause), MHD surface waves supported by two boundaries *together* can occur. It is shown that SW can be eigenmodes of the Earth's outer magnetosphere [17-22]. Their wavelengths can exceed the (large-scale) plasma density gradients existing at the plasmapause and magnetopause boundaries and exceeds also, the distance between the boundaries (of several Earth's radii $R_E$). The SW can propagate along adjacent plasma boundaries. If the two boundaries are the magnetopause and the plasmapause, the SW field simultaneously penetrates in the outer magnetosphere, magnetosheath, plasmasphere and ionosphere [19]. The SW energy density fluxes ought to be directed along the background magnetic field surfaces either azimuthally and/or to the Earth ionosphere where they are reflected or dissipated there. A formation of SW standing wave structures along the magnetic field fluxes is thus enabled, hence, this would result in a set of discrete eigenfrequencies (surface wave resonances, SWR). Their frequencies naturally ought to be dependent on the length of the magnetic field flux structure (shell). Such SWR can be easily driven by forces that disturb the questioned magnetic flux structure through its boundaries. This means that any solar wind velocity or pressure changes impacting the Earth's magnetopause can directly drive such SW resonances.

Let us briefly recall the basic properties of surface waves supported by two boundaries of plasma slab. Such slab geometry is a rough approximation to the outer Earth's magnetosphere, however this geometry is capable to reveal the main characteristics of the SW modes supported by more complicated plasma boundaries. The outer magnetosphere is thus treated as a shell (or slab) of thickness $d$ confined by the two boundaries − plasmapause and the magnetopause. The sausage (symmetric) and kink (asymmetric) modes are the well-known surface wave eigenmodes of a plasma slab [17,23,24]. As a result the slab boundaries are moving either in phase or in anti-phase (in the latter case the motion becomes as flapping one).

A little known fact is that a fully emptied slab surrounded by plasma also supports sausage and kink SW modes. Let us assume an empty slab/layer of thickness $d$ within plasma submerged in a constant magnetic field $B_o$. The magnetic field lines are aligned along the two plasma-slab boundaries, i.e. the plasma boundaries represent magnetic field line surfaces. A frame of reference is chosen so that $z$ axis is along the magnetic field $B_o$, $x$ axis − perpendicular to the slab boundaries, while $y$ axis completes the coordinate system. This means that $y$ and $z$ axes are parallel to the plasma-slab boundaries. According to normal mode analysis we look for surface modes of the following form

$$\{A_1\exp(-kx)+A_2\exp[k(x-d)]\}\exp(ik_z z + ik_y y - i\omega t), \quad 0 < x < d \quad \text{(the slab)}$$

$$A_\pm \exp(\pm\kappa x)\exp(ik_z z + ik_y y - i\omega t), \quad \text{(the plasma)}$$

where $A_{1,2}$ and $A_\pm$ are amplitudes to be determined by the boundary conditions at $x = 0$ and $d$, where $d$ is the slab thickness; $k$ and $\kappa$ are attenuation coefficients characterizing the attenuation of the surface mode field away from the boundary surfaces; Note that in magnetospheric physics context  $x$ coordinate corresponds to the dipole coordinate ν); $k_z$, $k_y$ are wavenumbers (where $k_z$ corresponds to dipole coordinate μ, and $k_y$ – to azimuthal number $m/R_E$ ); $\omega$ is the frequency of the surface modes; sign + refers to region $x < 0$ , and sign −: for region $x > d$. It can be easily demonstrated that the field attenuation coefficient $k$ of the MHD surface wave field in the empty slab (i.e. for $0 < x < d$) exactly coincides with the total wavenumber $k \equiv \sqrt{k_z^2 + k_y^2}$ [18]. Attenuation coefficient of the surface mode field in plasma environments $\kappa$ however is expected be different from total wavenumber $k$; $\kappa$ usually is frequency dependent and can vary between 0 and the total wavenumber $k$ [19]. The case of interest is, if the slab thickness $d$ is comparable to, or less than the inverse of the total wavenumber $k$, i.e. when $d \sim 2\pi/k$, or $kd < 1$. As the argument ($kx$) is small (because $kd < 1$), then $\exp(kd) \cong 1+kx$ and the surface mode field within the slab ( $|x| < d$) varies *linearly* in $x$, i.e. with a linear coefficient equal to the total wavenumber $k$. This suggests that surface mode field of the slab can behave roughly as a mode with a non-vanishing amplitude in $x$ (having in mind that $x$ corresponds to radial coordinate or latitude). Obviously, such a situation always happens when the surface mode wavelength exceeds the slab thickness $d$! Hence, for modes whose wavelengths are sufficiently great, the surface mode amplitude can stay approximately at its maximum value for a wide region. Consequently, the surface mode field can weakly attenuate in $x$ in the slab (through all the outer magnetosphere). In general, the MHD surface mode field possesses amplitude maxima at the two boundaries and goes through a minimum at the middle of the slab. Further, Nenovski and Momchilov [18] found that the MHD surface modes require at least two (magnetic) field components: along and perpendicular to the slab. They are also elliptically polarized, and the polarization of a slab changes its sign of polarization three times: at the two boundaries the middle of the slab.

   Let us apply these findings to the Earth's magnetosphere. The plasma density of the outer magnetosphere is typically only of one order less than the plasma densities in the magnetosheath and of several orders less than the plasmasphere densities [25]. The outer magnetosphere thus might be approximated as a rarefied or empty plasma slab, bounded by the magnetopause and plasmapause (being rich of plasma). According to surface wave theory, this structure admits propagation of MHD surface modes (sausage and kink) that can propagate along the boundary surfaces (plasmapause and magnetopause) both along the magnetic field and/or azimuthally. Further, their field entirely crosses the outer magnetosphere and penetrates both in the plasmasphere (deeply) and beyond the magnetopause (in the attached magnetosheath). field lines. A theory of global SW modes of the earth's magnetosphere has been conducted in Nenovski et al's paper [19].

   Further, the global SW mode frequency can become quantized. Once excited, the global SW mode will propagate along the boundary surfaces and can be bounced at some reflecting points, e.g. the ionosphere. The SW mode can thus bounce between the ionospheres of the Northern and Southern hemispheres and so SW resonance (SWR) can emerge. Then, the SWR frequency ω (being dependent on parallel wavelength $k_z$) has to be determined by the magnetic field line length and hence, quantized: $\omega_n = f(k_{z,n})$ where $k_{z,n} \sim \pi n/L$, where n is an integer and $L$ representing the average (to be determined) length of the magnetic slab (shell).

**Similarity and differences between SW resonances and field-line resonances (FLR).** As it was underlined the low frequency branch of the SW modes possesses discrete frequencies in the 1-4 mHz band. Global SW modes can be observed simultaneously at high and mid latitudes. Hence, the SWRs can easily be misinterpreted as cavity modes. Then, it is very important to distinguish SWR from the cavity ones. There are at least three ways to do it. First, a study of the mode frequency behavior under various magnetospheric conditions is requested. Second, the

SWR exercises polarization change at its amplitude extrema – a feature known for the FLR. Third, the SWR field easily penetrates beyond the region of localization of the cavity modes, the former could be observed both at high latitudes (at open field lines), and at low latitudes – deeply in the plasmasphere.

On the other hand, the SWR propagate to the Earth in manner similar to field-line resonances (FLR). Hence, it is important to point out also the differences between the global SWR and the FLR. The principal difference is that SWR are compressional, while the FLR are shear modes, i.e they possess different field polarization properties. Further, the SWR frequencies are sensitive to various plasma and geometry parameters (e.g. plasma density, different positions of the plasmapause (in LT), etc.). Hence, both the plasma density, the magnetic field magnitudes at the boundaries, as well as geometry parameter distribution will control the possible localisation of the SWR field structure in azimuth, i.e. the magnitude of $k_\perp$. Exact SWR field structure depends also on the plasma density radial distribution in the three regions. Below we classify the most of SWR and FLR similarities and differences:

♦ The SWRs the outer magnetosphere propagate to and back the Earth ionosphere forming standing field structures. The SWR field penetrates in all three directions: radially, azimuthally and meridionally, i.e. their fields can be observed both in the magnetosphere and in the ionosphere (high to low latitudes);

♦ The low frequency branch of the SWR possesses discrete frequencies in the 1-4 mHz band. Cavity modes of frequencies in the 1-4 mHz range practically do not exist unless an unlikely quarter-wavelength wave structure is admitted;

♦ The SWR model suggests that the discrete modes of very low frequencies (in the mHz band) can be observed both at high and mid and low latitudes. (Low-frequency FLR cannot however exist at mid and low latitudes.);

♦ The low frequency branch of the SWR retains its eigenfrequencies comparatively constant under variable solar wind density conditions. The SWR frequency depends on the magnetic line lengths of the outer Earth's magnetosphere. On the contrary, the cavity mode eigenfrequencies however are dependent both on the solar wind speed and density variations that crucially influence the magnetopause position;

♦ The SWRs possess polarization reversal across the amplitude maximum – a polarization property similar to the FLR. Hence, authentic SWR events in the magnetosphere might easily be misinterpreted as FLRs;

♦ A polarization reversal of the SWRs is expected at their field amplitude minimum expected between the magnetopause and plasmapause baundaries;

♦ Field amplitude peaks of the SWR are expected to be located simultaneously at high latitudes (close to the auroral boundary) and mid latitudes (close to plasmapause position);

♦ The SWR field amplitude distribution reaches its maximum at the plasma gradients and its structure is distributed *over* the whole gradient region, i.e. the SWR field distribution area have to encounter adjacent regions – plasmasphere and plasmatrough. (The FLR field structure emerges however *within* a region of strong plasma density gradients, e.g. in the plasmapause latitudes only).

♦ SWRs properly are resonances of the Earth's magnetosphere. Hence, SWRs may be driven/enhanced (as FLR itself) under conditions such as storm times, etc. Once driven, SWRs can persist for long time, e.g. during the storm recovery phase, because the magnetic field lines

(closed ones) in the outer Earth's magnetosphere practically even curved retain their length due to magnetic field flux concervation principle.

**Surface wave mode vs cavity modes.** Finally, comparing the SWR properties with the cavity modes, it is worth noting that (i) the SWR discrete frequencies appear to be definitely lower than those of the cavity modes: the cavity mode theory assumes that the cavity mode frequency is inversely proportional to the size $\Delta R$ of the outer magnetosphere, i.e. the radial distance between the plasmapause and the magnetopause (or bow shock) shortened under high speed and or high density solar wind. This distance is much less (at least two-three times) than the length of the magnetic field shell (slab) of the outer Earth's magnetosphere (its dayside or flank magnetosphere part); (ii) the SWRs and their discrete frequencies are expected to be more stable than the cavity mode frequencies. The reason is that the radial size of the magnetosphere strongly varies with magnetospheric conditions, while the magnetic field lines lengths do not change considerably with magnetospheric conditions. Thus, the observed characteristic frequencies of the discrete modes are not consistent with cavity/waveguide resonant frequencies.

**Conclusion.** The wave properties of the global SW modes (that follow from the general SW theory) suggest that the SWRs (especially the low frequency branch) might reconcile most of the properties of the observed discrete global Pc5 pulsations. Observational facts enlighten by Villante et al [8,9], Nenovski et al. [19] and Marin et al [10], favor the interpretation that (MHD) surface wave modes/resonances might be a real phenomenon in the Earth's magnetosphere.

The global SW modes therefore, should be recognized as main global resonances of the earth's magnetosphere that may be excited by various mechanisms (e.g. high speed solar wind). Global SWRs can emerge and what is more importantly, can survive for much long time compared to cavity/waveguide modes. Unlike the cavity/waveguide modes, the global surface wave resonances (SWR) occurring in the Earth's magnetosphere incorporate basic properties of the FLR and thus can straightforwardly be recorded on the ground.

REFERENCES

[1] Stewart B. On the great magnetic disturbance which extended from August 28 to September 7, 1859, as recorded by photography at the Kew Observatory, Philosophical Transactions of the Royal Society London, 1861, 423–430.
[2] Saito, T. Geomagnetic pulsations, Space Science Rev.,10, 1969, 319-412.
[3] Ruohoniemi J.M., R.A. Greenwald, K.B. Baker, J.C. Samson. HF radar observations of Pc5 field line resonances in the midnight/early morning MLT sector, J. Geophys. Res. 96, 1991, 15697-15710.
[4] Samson J.C., R.A. Greenwald, J.M. Ruohoniemi, T.J. Hughes D.D. Wallis.Magnetometer and radar obervtions of magnetohydrodynamic cavity in the erath's magnetosphere, Can. J. Phys. 69,1991, 929-931.
[5] Samson J.C., B.G. Harrold, J.M. Ruohoniemi, R.A. Greenwald, and A.D.M. Walker. Field line resonances associated with MHD waveguides in the magnetosphere, Geophys. Res. Letters, 19(5), 1992, 441-444.
[6] Provan G., T.K. Yeoman. A comparison of field-line resonances observed at the Goose Bay and Wick radars, Ann. Geophysicae 15, 1997, 231-235.
[7] Francia P., U. Villante. Some evidence for ground power enhancements at frequencies of global magnetospheric modes at low latitude, Ann Geophysicae, 15, 1997,17.

[8] Villante U., P. Francia, S. Lepidi, M. De Lauretis, E. Pietropaolo, L. Cafarella, A. Meloni, A.J. Lazarus, R.P. Lepping, F. Mariani. Geomagnetic field variations at low and high latitude during the January 10-11, 1997 magnetic cloud, Geophys.Res. Lett. 25, 1998, 2593-2596.

[9] Villante U., P. Francia, S. Lepidi. Pc5 geomagnetic field fluctuations at discrete frequencies at a low latitude station, Ann. Geophysicae, 19, 2001, 321-325.

[10] Marin J., V. Pilipenko, O. Kozyreva, M. Stepanova, M. Engebretson, P. Vega, E. Zesta.: Global Pc5 pulsations during strong magnetic storms:excitation mechanisms and equatorward expansion, Ann. Geophys., 32, 2014,1–13; doi:10.5194/angeo-32-1-2014.

[11] Kivelson M.G., J. Etcheto, J.G. Trotignon. Global compressibional oscillations of the terrestrial magnetosphere: The evidence and a cavity model. J. Geophys. Res. 89: doi: 10.1029/JA080i011p09851. issn: 0148-0227,1984.

[12] Kivelson M.G., D.J. Southwood. Resonant ULF waves: A new interpretation, Geophys. Res. Lett., 12(1), 1985, 49-52.

[13] Crowley G., W.J. Hughes, T.B. Jones, T.B.: Observational evidence of cavity modes in the Earth's magnetosphere, J. Geophys. Res.: Space Physics, 92, A11, 1987,12233–12240.

[14] Kivelson M.G., M. Cao, R.L. McPherron, R.J. Walker. A possible signature of magnetic cavity mode oscillations in ISEE spacecraft obseations, J. Geomagn. Electricity, 49, 1997, 1079-1098.

[15] Hartinger M., V. Angelopoulos, M.B. Moldwin, Y. Nishimura, D.L. Turner, K-H. Glassmeier, M.G. Kivelson, J. Matzka, J.,C. Stolle. Observations of a Pc5 global (cavity/waveguide) mode outside the plasmasphere by THEMIS, J. Geophys. Res., 117, 2012, p. A06202.

[16] Rae, I. J. and Donovan, E. F. and Mann, I. R. and Fenrich, F. R. and Watt, C. E. J. and Milling, D. K. and Lester, M. and Lavraud, B. and Wild, James A. and Singer, H. J. and Rème, H. and Balogh, A. (2005) Evolution and characteristics of global pc5 ULF waves during a high solar wind speed interval, J. Geophys. Res., 110 (A12211). pp. 1-16. ISSN 0148-0227; doi:10.1029/2005JA011007.

[17] Ballai I., R. Erdelyi, B. Roberts. Ducted compressional waves in the magnetosphere in the double-polytropic approximation, Annales Geophysicae. 01/2002; doi:10.5194/angeo-20-1553-2002.

[18] Nenovski P. G. Momchilov. Model of polarization state of compressional surface MHD waves in the low-altitude cusp, Planet. Space. Sci., 35, 1987,1561-1316.

[19] Nenovski P., U. Villante, P. Francia, M. Vellante, A. Bochev. Do we need a surface wave approach to the magnetospheric resonances? Planet. Space Sci., 55, 2007, 680-693.

[20] Nenovski P. Plasma sheet modes and their connection with Pi2 pulsations, Compt.rend.Acad.bulg.Sci., 31,1978, 1297-1300.

[21] Nenovski P.I. Plasma model of low-altitude cusp and polarization structure of compressional magnetohydrodynamical waves, Compt.rend.Acad.bulg.Sci. (Sofia), 38, No 10, 1985, 1331-4.

[22] Arshinkov I.S., A.Z. Bochev, D.L. Danov, P.I. Nenovski, On the interpretation of field-aligned currents in the magnetosphere, *Compt.rend.Acad.bulg.Sci*. (Sofia), 46, No 4, 1993, 53.

[23] Roberts B. Wave propagation in a magnetically structured atmosphere I: Surface waves at a magnetic interface, Sol. Phys 69, 1981a, 27-38.

[24] Roberts B. Wave propagation in a magnetically structured atmosphere II: Waves in a magnetic slab, Sol. Phys 69, 1981b, 39-56.

[25] Phan T.D., G. Paschmann, W. Baumjohann, N. Sckopke (1994), The magnetosheath region adjacent to the dayside magnetopause: AMPTE/IRM observations, J. Geophys. Res., 99, 121-141.